\documentclass[%
reprint,
superscriptaddress,
 amsmath,amssymb,
pra,
]{revtex4-2}

\usepackage{graphicx}% Include figure files
\usepackage{dcolumn}% Align table columns on decimal point
\usepackage{bm}% bold math
\usepackage{amsmath}
\usepackage{amssymb}
\usepackage[utf8]{inputenc}
\usepackage[T1]{fontenc}
\usepackage{mathptmx}
\usepackage{xcolor}
\usepackage{enumitem}   
\usepackage{subfig}
\usepackage[format=plain,labelfont=up,textfont=up]{caption}
\usepackage{dsfont}
\usepackage{xr-hyper}
\usepackage{hyperref}
\usepackage{url}
\usepackage{siunitx}
\usepackage{float}
\usepackage{soul,xcolor}

\usepackage{textgreek}
\usepackage{upgreek}

\begin{document}

\title{Surface Second Harmonic Generation in Centrosymmetric Molecular Crystalline Materials: How Thick is the Surface?}

\author{Benedikt Zerulla*}
\affiliation{Institute of Nanotechnology,
Karlsruhe Institute of Technology (KIT),
Kaiserstr. 12, 76131 Karlsruhe, Germany}
\email{benedikt.zerulla@kit.edu\\carsten.rockstuhl@kit.edu\\ivan.fernandez-corbaton@kit.edu\\marjan.krstic@kit.edu}
\author{Alejandro Luna D\'iaz}
\affiliation{Institute of Theoretical Solid State Physics,
Karlsruhe Institute of Technology (KIT),
Kaiserstr. 12, 76131 Karlsruhe, Germany}

\author{Christof Holzer}
\affiliation{Institute of Theoretical Solid State Physics,
Karlsruhe Institute of Technology (KIT),
Kaiserstr. 12, 76131 Karlsruhe, Germany}

\author{Carsten Rockstuhl*}
\affiliation{Institute of Nanotechnology,
Karlsruhe Institute of Technology (KIT),
Kaiserstr. 12, 76131 Karlsruhe, Germany}
\affiliation{Institute of Theoretical Solid State Physics,
Karlsruhe Institute of Technology (KIT),
Kaiserstr. 12, 76131 Karlsruhe, Germany}

\email{carsten.rockstuhl@kit.edu}

\author{Ivan Fernandez-Corbaton*}
\affiliation{Institute of Nanotechnology,
Karlsruhe Institute of Technology (KIT),
Kaiserstr. 12, 76131 Karlsruhe, Germany}
\email{ivan.fernandez-corbaton@kit.edu}

\author{Marjan Krstić*}
\affiliation{Institute of Theoretical Solid State Physics,
Karlsruhe Institute of Technology (KIT),
Kaiserstr. 12, 76131 Karlsruhe, Germany}
\email{marjan.krstic@kit.edu}


\begin{abstract}
Second harmonic generation (SHG) is forbidden in centrosymmetric molecular materials. However, a signal is frequently observed from interfaces where the symmetry is broken. Whereas the effect can be phenomenologically accommodated, an \emph{ab initio} qualitative and quantitative description has remained elusive, preventing the exploration of fascinating questions such as how deep below the surface the second harmonic can still be generated. To answer such questions, we present an \emph{ab initio} multiscale approach to compute the total and layer-dependent intensity of surface SHG from molecular crystals. The microscopic origin of surface SHG is identified in layer-dependent models with embedding partial charges combined with density functional theory. The models show increasing symmetry-breaking distortions of the electron cloud around the molecules as the surface layer is approached. The SHG at the molecular level is determined using time-dependent density functional theory and then brought to the scale of macroscopic films through a rigorous self-consistent multiple scattering formalism capable of predicting the measurable optical intensities of the generated second harmonic signal. We study crystalline molecular films with centrosymmetric unit cells of 7,9-Dibromobenzo[\emph{h}]quinolin-10-ol. The intensity of the SHG at the surface layer is two orders of magnitude larger than at the next layer below and three orders of magnitude larger than two layers below. Besides providing fundamental understanding, our approach can be used for designing and optimizing optical devices containing nonlinear molecular materials, such as molecular  laminates. We show that a relatively basic Kretschmann-like setup can enhance the surface SHG of a crystalline film of centrosymmetric molecular unit cells a thousand times.
\end{abstract}

\maketitle

\section{Introduction}
%\section{}
Second harmonic generation (SHG) is a specific nonlinear light-matter interaction process far from the resonant frequencies of material transitions. It was observed for the first time by Franken \emph{et al.}\cite{PhysRevLett.7.118} in 1961 after the invention of the pulsed ruby optical laser \cite{MAIMAN1960,PhysRevLett.5.303}. In SHG, two incident photons of the same energy are combined into one photon of twice the energy, thus the name second harmonic generation. Its existence was theorized by Bloembergen \emph{et al.} \cite{PhysRev.128.606}, one year after they presented the solutions to Maxwell's equations in nonlinear dielectrics. Their theory states that an SHG signal from electric dipoles and all other even-order nonlinear optical processes is forbidden within centrosymmetric materials. At the same time, an SHG signal at the surface of the material was surprisingly reported from the centrosymmetric calcite crystal \cite{PhysRevLett.8.404}. In 1968, in another publication from Bloembergen and coworkers, it was undoubtedly theoretically proven that the signal is produced from the surface \cite{PhysRev.174.813}. The intuitive explanation is that SHG at a surface is allowed because the interface between two media breaks the inversion symmetry of the bulk. 

The experimentally observed surface SHG can be accommodated in Maxwell's equations through a second-order surface tensor at the material interfaces \cite{Dadap1999}. However, the tensor elements must typically be determined experimentally, and qualitative and quantitative explanations on a microscopic or molecular level are missing so far, preventing the \emph{ab initio} numerical study of surface SHG. In particular, the question of how deep into the material the surface SHG signal is still generated remains unanswered. 

In this paper, we compute the total and the layer-dependent intensity of surface SHG from molecular crystals. The methodology starts \emph{ab initio} at the molecular level, where the layer-dependent first hyperpolarizabilities are computed using embedded models in density functional theory (DFT) calculations. Such hyperpolarizabilities are then used in an optical scattering code to go from the molecular to the macroscopic scale and ultimately predict measurable SHG intensities from devices such as molecular films on substrates. Since the bulk SHG can be computed separately by similar means, our methodology can predict the contributions of bulk and surface terms to the total signal measurable in experiments.

In the rest of the article, we first demonstrate the validity of our approach by computing the bulk and surface SHG of the non-centrosymmetric molecular urea crystal and obtaining an excellent match to experimental results. That is done because the material is well understood, and we can clearly outline our numerical workflow. Then, we consider a thin crystal film made from the centrosymmetric 7,9-Dibromobenzo[h]quinolin-1. Generally, a bulk contribution to the SHG is not expected. However, the symmetry is broken at the interface, as evident by tiny deformations of the electron density of the molecules close to the surface. This gentle symmetry break allows the surface SHG. We compute the SHG signal strength from a unit cell inside each layer below the surface and observe that it rapidly decays as the distance from the surface increases. The intensity of the SHG at the surface layer is two orders of magnitude larger than at the next layer below and three orders of magnitude larger than two layers below. The interlayer distance, in this case, is 1.57303\,nm. In the next step, we use a multiscale methodology \cite{first_paper} to predict SHG field strengths emitted from thin films of crystalline 7,9-Dibromobenzo[h]quinolin-1 as a function of their film thickness up to 1.5\,$\upmu$m, and show variations due to Fabry-Perot resonances of the planar films. As expected, the rapid decay of the SHG toward the inside of the film is also observed at the macroscopic scale. Finally, we design a Kretschmann-like experimental setup for enhancing the surface SHG by illumination with the surface evanescent wave produced by total internal reflection. The scheme enhances the surface SHG by a factor of $\approx$1000. 

\section{Results and Discussion}

\subsection{The Proof of Principle on the Example of an Embedded Surface Model of Urea}

The inspiring pioneering work of C. L. Tang and collaborators \cite{CASSIDY1979243, HALBOUT1979, Morrell1979} in preparation of high-quality single crystals of urea and consequent experimental investigations of SHG combined with complementary measurements and theoretical work of J. Zyss and coworkers \cite{10.1063/1.444266, LEDOUX1982203} motivated us to use this molecular material in the first part of the manuscript as a molecular model to demonstrate the proof of principle of our approach. We focus on the single unit cell of urea embedded into the field of calculated partial charges of the same unit cell. This demonstrates that studying the second harmonic generation within the bulk of the material and at the surface is possible. The material is chosen because a larger number of reference results exists, both from a computational and experimental perspective. 

\begin{figure*}
\centering
\includegraphics[width=1.0\textwidth,keepaspectratio]{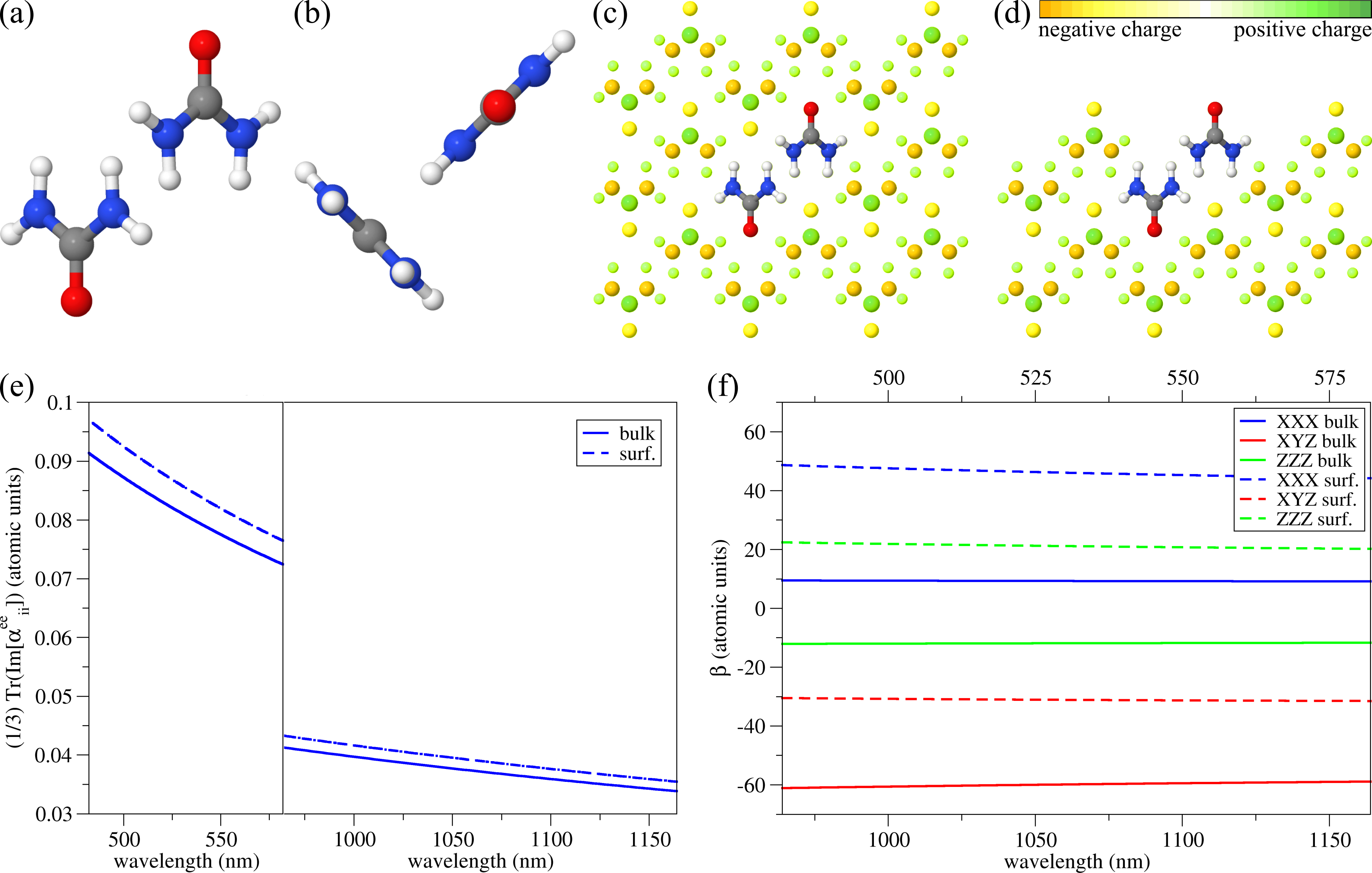}
 \hspace{0cm}
  
  \caption{\textbf{(a)} The side-view of the unit cell of a urea crystal composed of two urea molecules. \textbf{(b)} The top-view of the same unit cell of urea. White, grey, blue, and red spheres represent hydrogen, carbon, nitrogen, and oxygen atoms, respectively. \textbf{(c)} The side-view of the embedding of the urea unit cell into the surrounding field of partial charges representing the bulk model. The charge field covers the size of 3x3x3 unit cells. Partial charges are represented with discrete spheres. The yellow and green colors represent negative and positive partial charges scaled relatively from the lowest negative to the highest positive value. The value of each charge represents the intensity of the color. The partial charges have been removed in the center of the partial charges' field, where the explicitly treated unit cell is located. \textbf{(d)} The side-view of the surface model of the unit cell of urea embedded into the 3x3x2 unit cells' large partial charge field representing the unit cell of the urea crystal at the surface of the material. The partial charges were removed in the center of the surface, and the explicitly treated unit cell of the urea was placed there. \textbf{(e)} Absorption of urea calculated from electric-electric dynamic polarizabilities in atomic units.\textbf{(f)} The most significant components of the real part of the complex first hyperpolarizability tensors in the range 964-1164\,nm in atomic units.}
  \label{fgr:urea}
\end{figure*}

The unit cell of urea was obtained from the crystallographic data file published by Guth \emph{et al.} and visualized in Figures~\ref{fgr:urea}\textbf{(a)} and \textbf{(b)} \cite{GuthHegerKleinTreutmannScheringer}. The region we treat quantum-chemically in our models to calculate the molecular properties consists of two molecules of urea oriented as in the unit cell. The surrounding of this unit cell was considered at the level of partial charges that replace the atoms in the other unit cells. 3x3x3 unit cells of partial charges were considered for the bulk. Partial charges in the middle of the unit cell containing the explicitly considered quantum region were removed, of course. For the surface model, we constructed the field of partial charges with the size of 3x3x2 unit cells and removed the charges from the middle of the surface to insert the quantum region. Both models are visualized in Figures~\ref{fgr:urea}\textbf{(c)} and \textbf{(d)}. The linear and nonlinear properties of these molecules in the bulk and on the surface are computed with techniques described in the Methods section. 
The linear and nonlinear molecular properties for the bulk and the surface model are shown in Figures~\ref{fgr:urea}\textbf{(e)} and \textbf{(f)}, respectively. The results clearly indicate that embedding the molecules into a partial charge field of the surrounding drastically influences the nonlinear optical properties. While the linear polarizabilities ($\alpha$) increase only mildly for the surface model, the calculated first hyperpolarizabilities ($\beta$) increase substantially for the surface compared to the bulk model. 

We then applied the approach from Miniewicz \emph{et al.}\cite{Miniewicz2019} to calculate the macroscopic second-order susceptibility for the bulk and surface model of urea from the calculated dynamic polarizabilities and first hyperpolarizabilities. 
The predicted values for photons at an incident wavelength of 1064\,nm are $d_{14}^{\mathrm{(2),bulk}}=1/2\mathbf{\chi}_{14}^{\mathrm{(2),bulk}}=-1.72\,\mathrm{pm}\,\mathrm{V}^{-1}$ and $d_{14}^{\mathrm{(2),surface}}=1/2\mathbf{\chi}_{14}^{\mathrm{(2),surface}}=-0.91\,\mathrm{pm}\,\mathrm{V}^{-1}$. These computed values are fairly close to the measured values of $d_{14}^{\mathrm{(2),exp}}=1.2-1.4\,\mathrm{pm}\,\mathrm{V}^{-1}$ reported in the work of Miniewicz \emph{et al.} \cite{Miniewicz2019} and Levine \emph{et al.} \cite{Levine1993} The publication from Ferrero \emph{et al.} \cite{Ferrero2009} reports that the calculated values of the urea crystal are $d_{14}^{\mathrm{(2)}}=-0.908\,\mathrm{pm}\,\mathrm{V}^{-1}$, which matches the value for the surface model predicted here. The difference in sign in the reported values and calculations stems from the convention in the DFT implementation of the nonlinear optical (NLO) calculations and the overall orientation of the crystals in the experiments. These results gave us the confidence that the embedding of the quantum region into a field of partial charges can be used to study the emergence of surface SHG in molecular crystalline materials from first principle quantum chemistry simulations.   

\subsection{Embedded Models of Centrosymmetric 7,9-Dibromobenzo[\emph{h}]quinolin-10-ol Crystal from Bulk to the Surface}

The second part of this work focuses on a different molecular crystalline material. Here, we chose a centrosymmetric material built from 7,9-Dibromobenzo[\emph{h}]quinolin-10-ol molecules, a benzo[h]quinolin-10-ol derivative, with a monoclinic spacegroup P2\textsubscript{1}/n \cite{Tsai2017}. The centrosymmetry suggests a vanishing bulk first hyperpolarizability. A possible SHG signal in this material should exclusively come from the broken symmetry at the crystal surface. We will study the nonlinear response from thin films of the material that we divide into a finite number of layers. We will carefully study the emergence of that nonlinear response by faithfully describing the first hyperpolarizabilities in each layer of the molecular material using the quantum model. The presence of the other layers is considered at the level of the partial charges. By changing from a situation in bulk, where partial charges are symmetrically arranged around a given molecular layer, to a situation at the surface, where partial charges occupy only one half-space relative to the considered molecule layer, we expect a pronounced surface SHG signal to emerge, and we wish to quantify that effect. Most importantly, this allows us to pinpoint the spatial domain from which a noticeable nonlinear response emerges. 

\begin{figure*}
\centering
\includegraphics[width=0.75\textwidth,keepaspectratio]{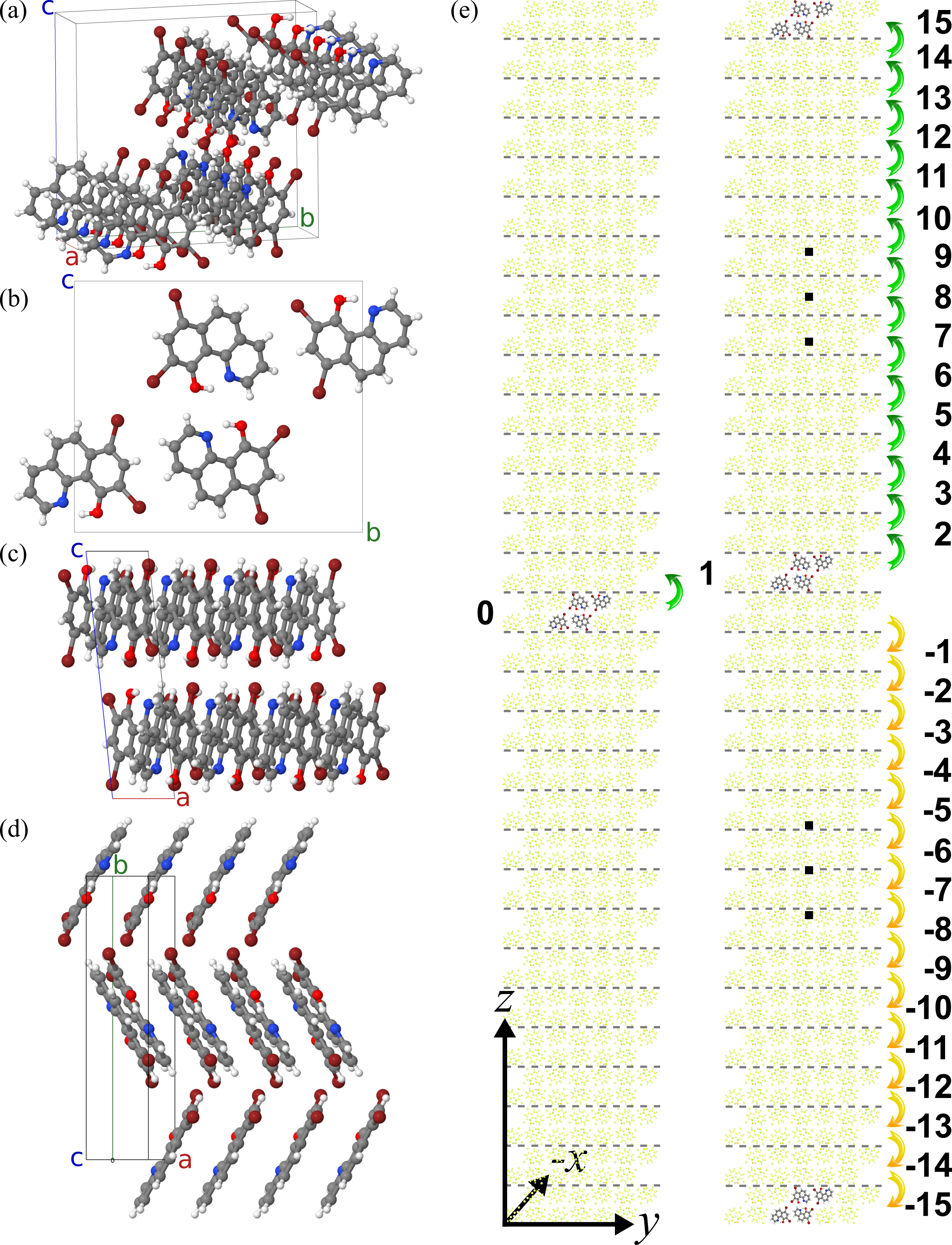}
 \hspace{0cm}
  
  \caption{\textbf{(a)} A quantum region consisting of four unit cells of 7,9-Dibromobenzo[\emph{h}]quinolin-10-ol crystal stacked along vector \emph{a}. \textbf{(b)} The view down the -\emph{a} vector. \textbf{(c)} The view down the +\emph{b} vector. \textbf{(d)} The top-down view. \textbf{(e)} A visualization of each of the 31 embedded models and placement of the quantum region in the field of the surrounding partial charges mimicking the periodic cell of the material as it is approaching the top (15) or bottom (-15) surface of the crystal. The model "0" represents the bulk surrounding in the middle of the material.}
  \label{fgr:layers}
\end{figure*}

The quantum region is chosen to be four unit cells of the crystal in the x-direction, visualized in Figures~\ref{fgr:layers}\textbf{(a)}-\textbf{(d)}. The unit cell consists of 384 atoms. To observe and understand the surface SHG, we again need to embed the quantum region into the field of the partial charges generated from all the other molecules forming the structure. We build a partial charge field by stacking 3x3x31 quantum regions along the translation vectors \emph{a}, \emph{b}, and \emph{c}, respectively. The last number indicates that we study a thin film consisting of 31 layers of the considered molecular crystal. To accommodate the quantum region, the partial charges were removed in their positions for each of the 31 models along the z-axis, yielding a total field of 106752 charges. The scheme of all models is presented in Figure~\ref{fgr:layers}\textbf{(e)}. The models are denoted by numbers from "-15" to "15", representing the bottom and top surface layers of the thin film of the material. Model "0" describes the situation where the considered molecule is located in the center of the partial charge field, representing the "bulk" case. Since the quantum region and the surrounding partial charge field are centrosymmetric, the calculated first hyperpolarizabilities are equal to zero, as expected. In the other models, built by moving the quantum region one vector \emph{c} length up or down within the charge field, the centrosymmetry is broken. Thus, the polarization of the surrounding on the quantum region is expected to produce an SHG signal in those models, yielding the highest values of $\beta$ on the top and bottom surface interface between the crystalline film and vacuum. There, the asymmetry is the strongest. Such behavior was indeed observed upon performing time-dependent density functional theory (TD-DFT) calculations of the first hyperpolarizabilities as presented in Table~\ref{tbl:beta} and Figure~S3 of the supplementary information (SI), for the most intensive components of the hyperpolarizability tensor. 

\begin{figure*}
\centering
\includegraphics[width=0.70\textwidth,keepaspectratio]{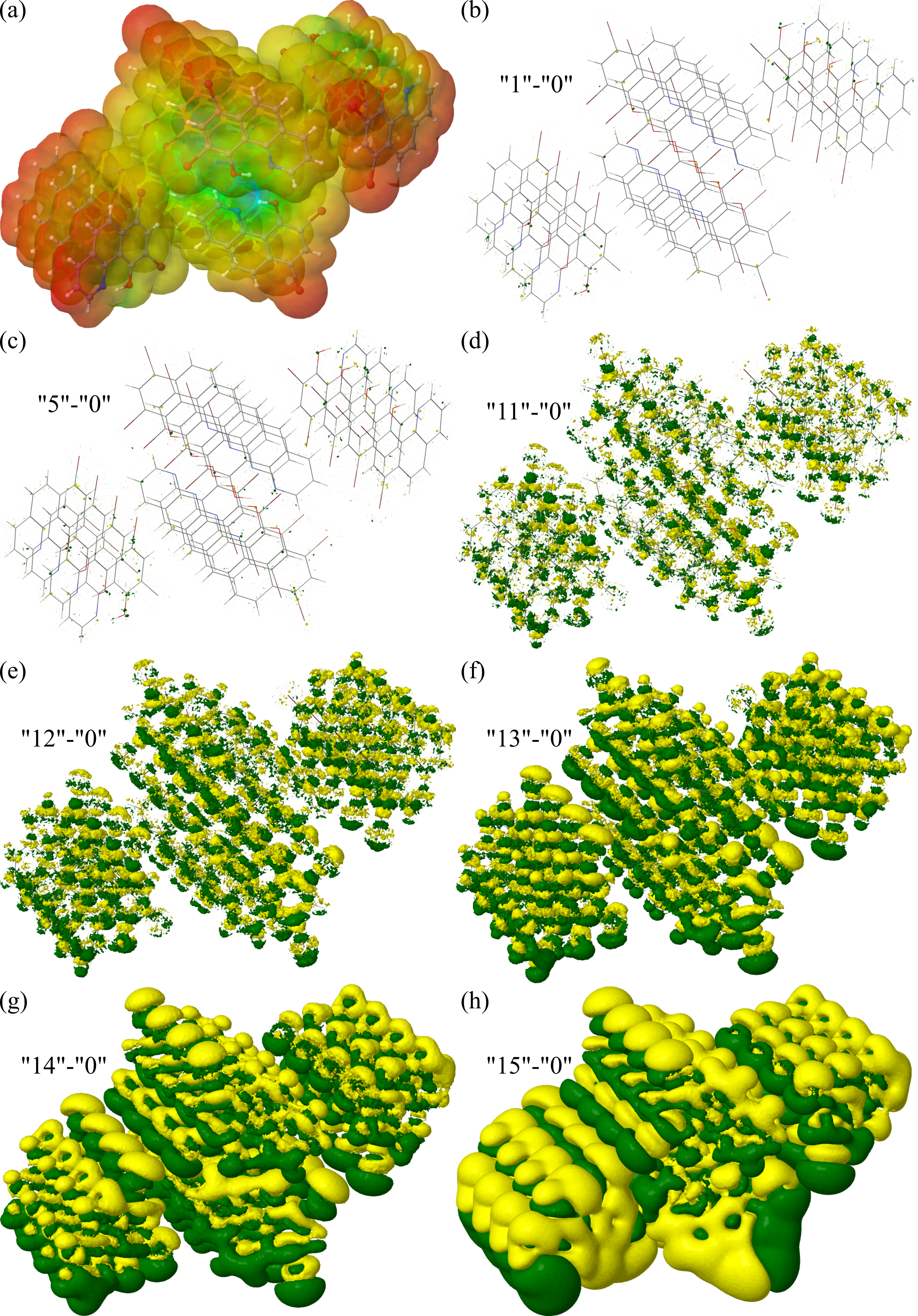}
 \hspace{0cm}
  
  \caption{\textbf{(a)} The electrostatic potential mapped on the total electron density. \textbf{(b-h)} The total electron density difference between models "1", "5", and "11"-"15" and model "0". Green and yellow color represent positive and negative differences in the total density, respectively. The isosurface is plotted for a cutoff value of \textpm 0.00001.}
  \label{fgr:dens_diff}
\end{figure*}

Furthermore, to explain the emergence of the surface SHG at the interface on the quantum level, we focused on the total electron density difference between models "15" to "1" and model "0". A possible difference visualizes the broken symmetry of the surroundings when the considered quantum region approaches the surface. The results are presented in Figure~\ref{fgr:dens_diff}. For the models "11"-"15" we observe a minuscule but physically significant change in the total electron density. In contrast, the molecules deeper under the surface experience the centrosymmetric surrounding. Therefore, they do not experience a change in the total electron density from the surrounding as depicted for the density differences "5"-"0" or "1"-"0" in Figure ~\ref{fgr:dens_diff}. In the depicted isosurfaces of the electron density difference, the cutoff was chosen to be \textpm 0.00001. The surrounding material below causes an additional polarization of the total electron density of molecules at the surface. Therefore, we conclude that the origin of the surface SHG stems from the influence of the material below the molecules located at the surface. In the molecule at stake, the electron density in the quantum region becomes slightly more positive at the top (+z direction) of the cell compared to the bulk model "0". In exchange, it gets slightly more negative at the bottom of the cell. In both situations, we observe a rise of a significant dipole moment (Table S3) within the molecules at the surface. Consequently, this causes an SHG signal from the centrosymmetric molecular crystalline material such as 7,9-Dibromobenzo[\emph{h}]quinolin-10-ol to emerge. The observed effect of minimal "breaking" of the centrosymmetry of the electron density in molecules is larger the closer the molecules to the surface(interface) in the material are. For convenience, all calculated total electron density differences are presented in Figures S4 and S5. 

Interestingly, matching top and bottom layers, e.g., "-15" and "15", have all the components of the first hyperpolarizability tensor of the same magnitude but with a sign reversal. This can be clearly seen from Table~\ref{tbl:beta}. This fact can be exploited in future studies to reduce the number of expensive TD-DFT calculations by half and still consider both surface interfaces in the simulated material on larger scales, obtaining further results faster and more energy efficient. The models of the periodic molecular materials deep inside the thin film ("-5" to "5"), around the "bulk" model "0", have almost negligible values of the hyperpolarizabilities. By tendency, we observe slight oscillations in the first hyperpolarizability tensor around zero. That oscillation can be attributed to numerical noise in the quadratic response calculations of such large embedded models. 

Having such detailed information about all models, we can further draw additional exciting conclusions about the emergence of the surface SHG signal in crystalline film materials. In Table~\ref{tbl:beta} in the brackets, we present percentage-wise intensities of each subsurface model > 0 compared to the surface mode "15" and all models < 0 to the surface model "-15" to easily grasp the overall build-up of the surface SHG signal. It is seen that after five unit cell vector lengths below the surface, the contribution to the overall signal drops below 1\%. Thus, one can conclude that, in principle, it is enough to consider only those first five surface layers in modeling the surface SHG in general Maxwell scattering multiscale simulations of crystalline materials and devices built thereof. A layer of that thickness is the spatial domain from which the SHG emerges. That corresponds to the spatial extent of the surface in the surface SHG. 

From this point on, we move away from the quantum chemistry calculations and analysis and use T-matrices and hyper-T-matrices constructed from TD-DFT results to study the observable nonlinear optical response of the stack of the molecular crystal in a multiscale fashion.

\begin{table*}
  \caption{The highest components of TD-DFT calculated damped first hyperpolarizability tensors for all models in atomic units rounded to the first decimal digit. In brackets, we provide information on the percentage values of each model compared to the value of the first hyperpolarizability of the surface model "15" for models > 0 and to model "-15" for models < 0. The incident photons have a wavelength of 1064~nm.}
  \label{tbl:beta}
  \small
  \begin{tabular}{ccccccc}
    \hline
     %& \multicolumn{6}{c}{$\beta$} \\%& \multicolumn{2}{c}{$\beta$} & \multicolumn{2}{c}{$\beta$}\\
     %\hline
     & \multicolumn{2}{c}{YYY}  & \multicolumn{2}{c}{YXY}  & \multicolumn{2}{c}{XXX} \\
     %\hline
    Model  & Re($\beta$) & Im($\beta$) & Re($\beta$) & Im($\beta$) & Re($\beta$) & Im($\beta$) \\
    
    \hline
15 &-2389.0(100.0) &-160.3(100.0) &-978.1(100.0) &-75.6(100.0) &-1050.7(100.0) &-86.1(100.0)\\
14 &-273.5(11.4) &-12.2(7.6) &-40.7(4.2) &-3.6(4.7) &-165.8(15.8) &-10.7(12.4)\\
13 &-115.7(4.8) &-5.1(3.2) &-14.9(1.5) &-1.4(1.8) &-67.9(6.5) &-4.4(5.1)\\
12 &-58.1(2.4) &-2.5(1.6) &-7.0(0.7) &-0.7(0.9) &-31.5(3.0) &-2.0(2.4)\\
11 &-32.0(1.3) &-1.4(0.9) &-3.5(0.4) &-0.3(0.4) &-16.1(1.5) &-1.0(1.2)\\
10 &-19.8(0.8) &-0.9(0.5) &-2.6(0.3) &-0.2(0.3) &-10.2(1.0) &-0.7(0.8)\\
9 &-12.1(0.5) &-0.5(0.3) &-1.4(0.1) &-0.1(0.2) &-6.4(0.6) &-0.4(0.5)\\
8 &-7.1(0.3) &-0.3(0.2) &-0.6(0.1) &-0.1(0.1) &-3.8(0.4) &-0.2(0.3)\\
7 &-5.8(0.2) &-0.3(0.2) &-1.0(0.1) &-0.1(0.1) &-3.3(0.3) &-0.2(0.3)\\
6 &-3.5(0.1) &-0.1(0.1) &-0.5(0.0) &0.0(0.1) &-2.3(0.2) &-0.2(0.2)\\
5 &-2.4(0.1) &-0.1(0.1) &-0.4(0.0) &0.0(0.0) &-1.7(0.2) &-0.1(0.1)\\
4 &-0.6(0.0) &0.0(0.0) &0.2(0.0) &0.0(0.0) &-0.9(0.1) &-0.1(0.1)\\
3 &-1.4(0.1) &-0.1(0.0) &-0.2(0.0) &0.0(0.0) &-0.7(0.1) &0.0(0.1)\\
2 &-0.2(0.0) &0.0(0.0) &0.1(0.0) &0.0(0.0) &-0.6(0.1) &0.0(0.0)\\
1 &0.2(0.0) &0.0(0.0) &0.1(0.0) &0.0(0.0) &-0.4(0.0) &0.0(0.0)\\
0 &0.0(0.0) &0.0(0.0) &0.0(0.0) &0.0(0.0) &0.0(0.0) &0.0(0.0)\\
-1 &-0.2(0.0) &0.0(0.0) &-0.1(0.0) &0.0(0.0) &0.4(0.0) &0.0(0.0)\\
-2 &0.2(0.0) &0.0(0.0) &-0.1(0.0) &0.0(0.0) &0.6(0.1) &0.0(0.0)\\
-3 &1.4(0.1) &0.1(0.0) &0.2(0.0) &0.0(0.0) &0.7(0.1) &0.0(0.1)\\
-4 &0.6(0.0) &0.0(0.0) &-0.2(0.0) &0.0(0.0) &0.9(0.1) &0.1(0.1)\\
-5 &2.4(0.1) &0.1(0.1) &0.4(0.0) &0.0(0.0) &1.7(0.2) &0.1(0.1)\\
-6 &3.5(0.1) &0.1(0.1) &0.5(0.0) &0.0(0.1) &2.3(0.2) &0.2(0.2)\\
-7 &5.8(0.2) &0.3(0.2) &1.0(0.1) &0.1(0.1) &3.3(0.3) &0.2(0.3)\\
-8 &7.1(0.3) &0.3(0.2) &0.6(0.1) &0.1(0.1) &3.8(0.4) &0.2(0.3)\\
-9 &12.1(0.5) &0.5(0.3) &1.4(0.1) &0.1(0.2) &6.4(0.6) &0.4(0.5)\\
-10 &19.8(0.8) &0.9(0.5) &2.6(0.3) &0.2(0.3) &10.2(1.0) &0.7(0.8)\\
-11 &32.0(1.3) &1.4(0.9) &3.5(0.4) &0.3(0.4) &16.1(1.5) &1.0(1.2)\\
-12 &58.1(2.4) &2.5(1.6) &7.0(0.7) &0.7(0.9) &31.5(3.0) &2.0(2.4)\\
-13 &115.7(4.8) &5.1(3.2) &14.9(1.5) &1.4(1.8) &67.9(6.5) &4.4(5.1)\\
-14 &273.5(11.4) &12.2(7.6) &40.7(4.2) &3.6(4.7) &165.8(15.8) &10.7(12.4)\\
-15 &2389.0(100.0) &160.3(100.0) &978.1(100.0) &75.6(100.0) &1050.7(100.0) &86.1(100.0)\\

    \hline
  \end{tabular}
\end{table*}

\subsection{Nonlinear Optical Response of Film of Molecular Crystal}
Whereas in the previous section, we studied the actual nonlinear material properties of the molecular crystal, we continue in this next step to consider the nonlinear optical response of the molecular crystal. The considered structure is shown in Figure~\ref{fgr:layers}\textbf{(e)}, and we consider it infinitely extended in the xy-plane in a vacuum. 

The nonlinear optical response is computed with the Hyper-T-matrix-based approach described in \cite{first_paper}. For this purpose, the Hyper-T-matrix is computed from the first hyperpolarizability tensor for each layer individually, arranged periodically in the xy-plane, and stacked in the z-direction. In the following, we also vary the number of layers denoted with "0" in Figure~\ref{fgr:layers}\textbf{(e)} that forms the bulk. Due to the concept of the approach from \cite{first_paper}, we can compute the SHG field originating from each layer individually. 

\begin{figure*}[]
\centering
     \subfloat{
	\includegraphics[width=0.47\textwidth]{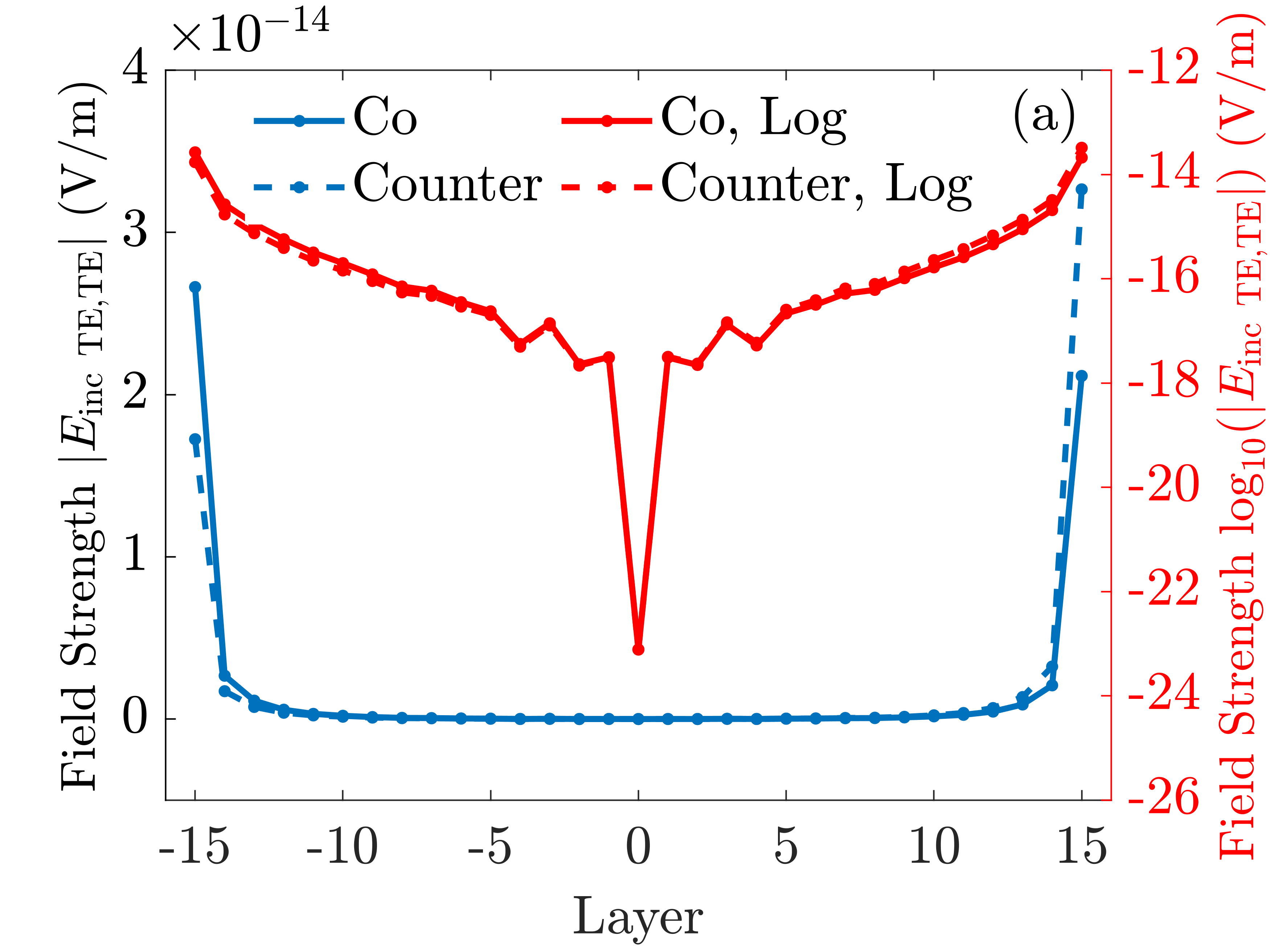}
 }\hspace{0cm}
 \subfloat{
	 \includegraphics[width=0.47\textwidth]{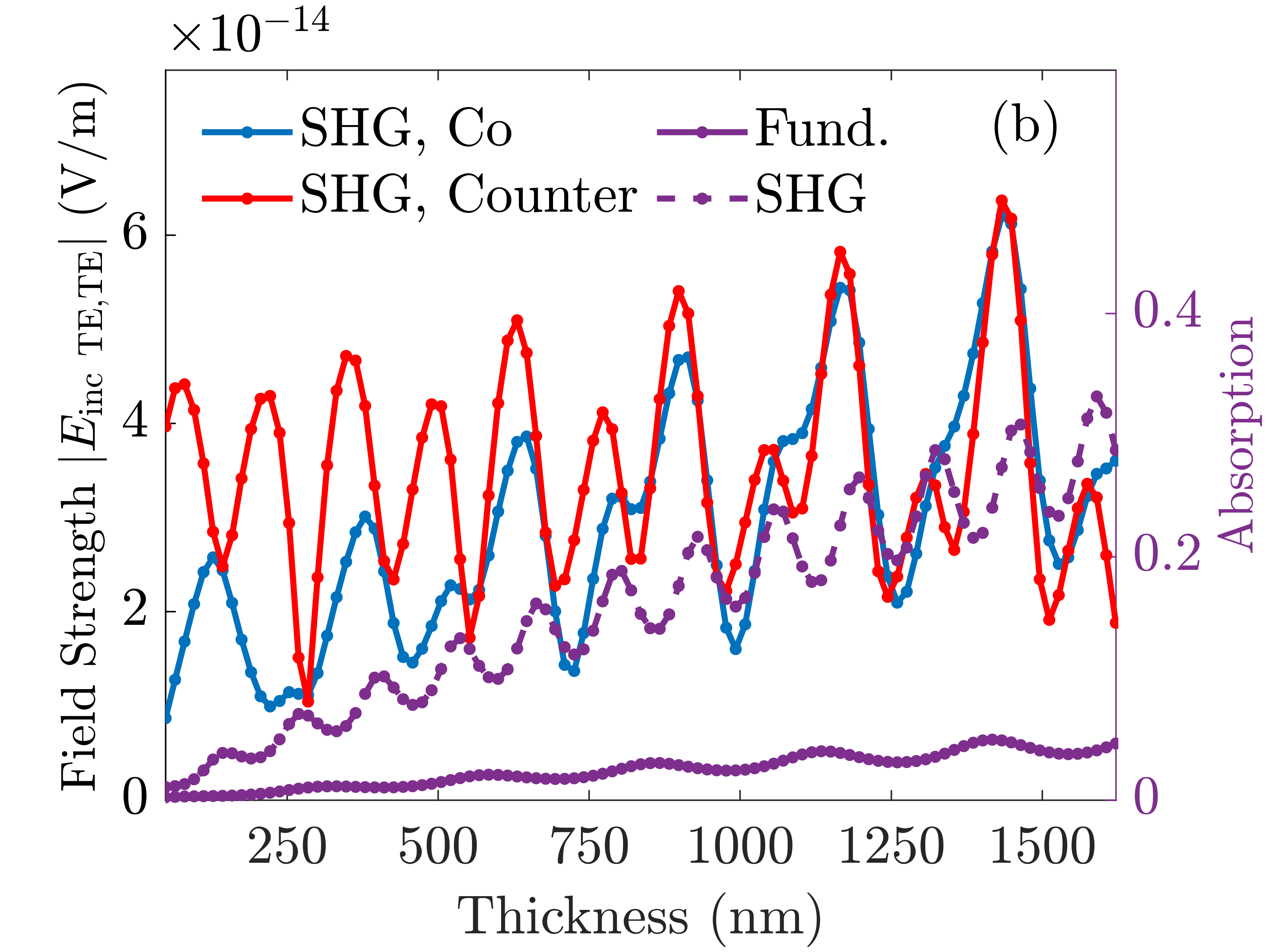}
}\\

	\caption{\textbf{(a)} Amplitudes of the SHG fields from the different layers of a molecular film consisting of the structure in Figure~\ref{fgr:layers}\textbf{(e)} infinitely extended in the xy-plane. The fields are co- and counterpropagating to the illumination. The structure is illuminated from below with a TE-polarized plane wave under normal incidence. The wavelength of the fundamental wave is 1064\,nm. The SHG fields of the bottom ("$-15$") and top ("$15$") surface layers are considerably stronger than the fields of any other layer. The contribution from the bulk layer ("$0$") is only because of numerical noise and is several orders of magnitude below those of other layers. \textbf{(b)} SHG fields from the film together with the linear absorption for a plane wave at the fundamental and the SHG frequency for different values of the film thickness, i.e., different numbers of bulk layers "0". One observes Fabry-Perot interferences for all quantities since all waves reflect off the interfaces between the film and vacuum and that, in general, the total SHG signal increases with the linear absorption.}
    \label{fig:LayersNormalIncidence}
	\end{figure*}

In Figure~\ref{fig:LayersNormalIncidence}\textbf{(a)}, the amplitudes of the SHG fields as generated from the individual layers are shown, which are co- and counterpropagating to the illuminating plane wave. Here, we set the number of bulk layers to one, i.e., we continue to consider the 31 layers previously discussed. In this simulation, we considered a TE-polarized normally incident plane wave illuminating the structure from below. The wavelength of the fundamental wave is 1064\,nm throughout the article. This TE-polarized plane wave suggests a polarization along the y-axis. The field amplitude of the illumination is in the following always set to $1\,\mathrm{V/m}$.
We observe in Figure~\ref{fig:LayersNormalIncidence}\textbf{(a)} that the SHG fields originating from the bottom and top surface layers are considerably larger than the contributions from the other layers. Approaching the bulk, the contributions to the fields decrease rapidly by several orders of magnitude. This is best appreciated in the logarithmic scale.
In Figure~\ref{fig:LayersNormalIncidence}\textbf{(b)}, the amplitudes of the SHG waves propagating in reflection (Counter) and transmission (Co) after the plane wave illumination and the linear absorption for the fundamental field and a field at the SHG frequency are depicted for different values of the thickness of the stack, corresponding to different numbers of bulk layers. All quantities show Fabry-Perot resonances as both the SHG and the fundamental fields are reflected off the interfaces between the molecular film and vacuum. 

\begin{figure*}[]
\centering
 \subfloat{
	\includegraphics[width=0.25\textwidth]{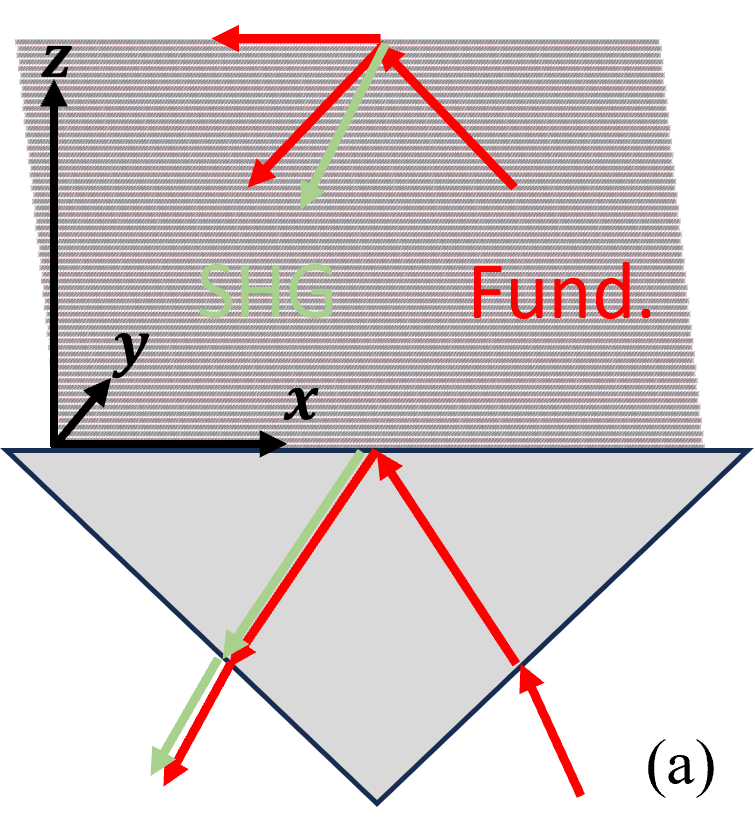}
 }
     \subfloat{
	\includegraphics[width=0.33\textwidth]{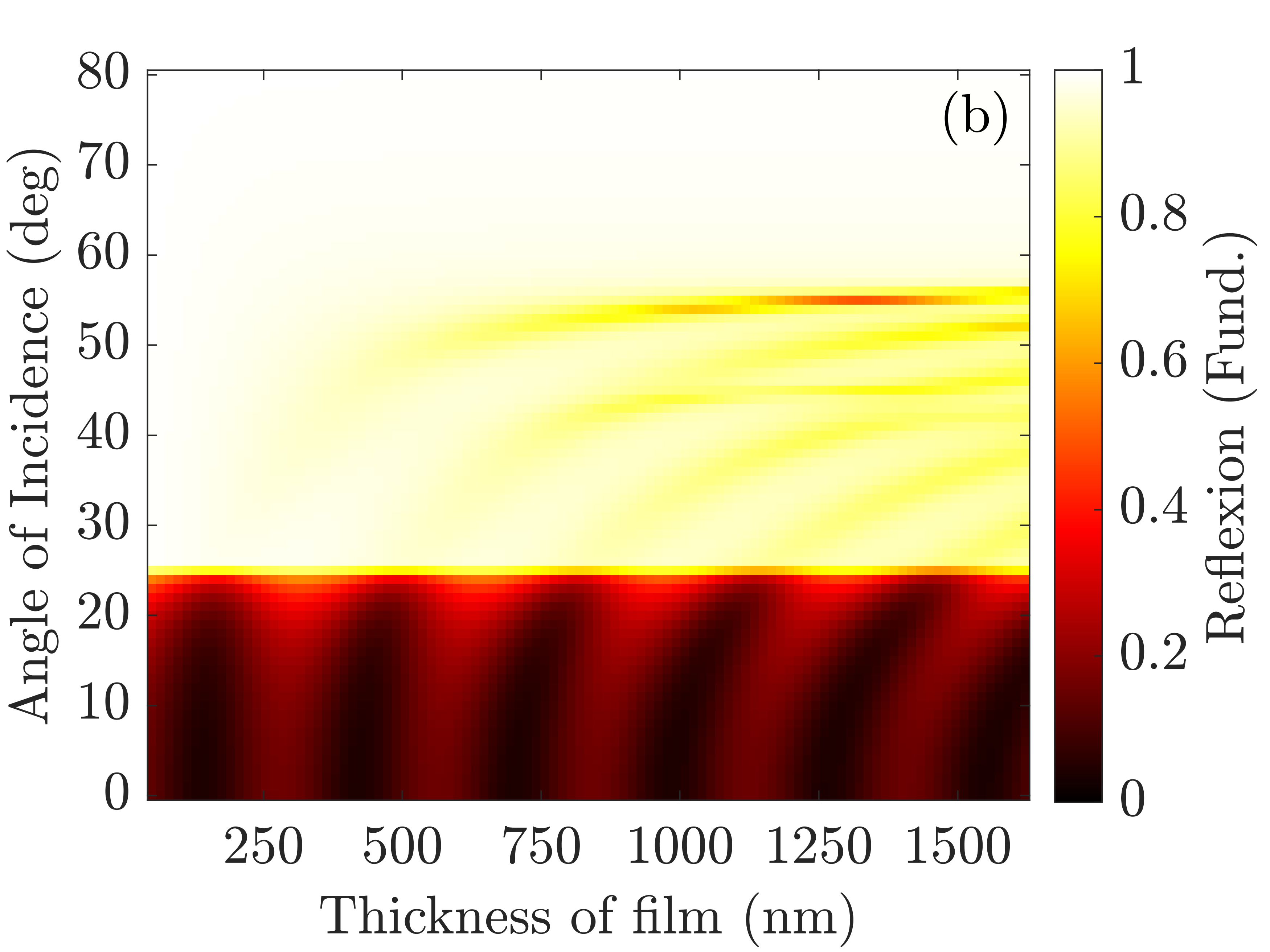}
 }\hspace{0cm}
 \subfloat{
	 \includegraphics[width=0.33\textwidth]{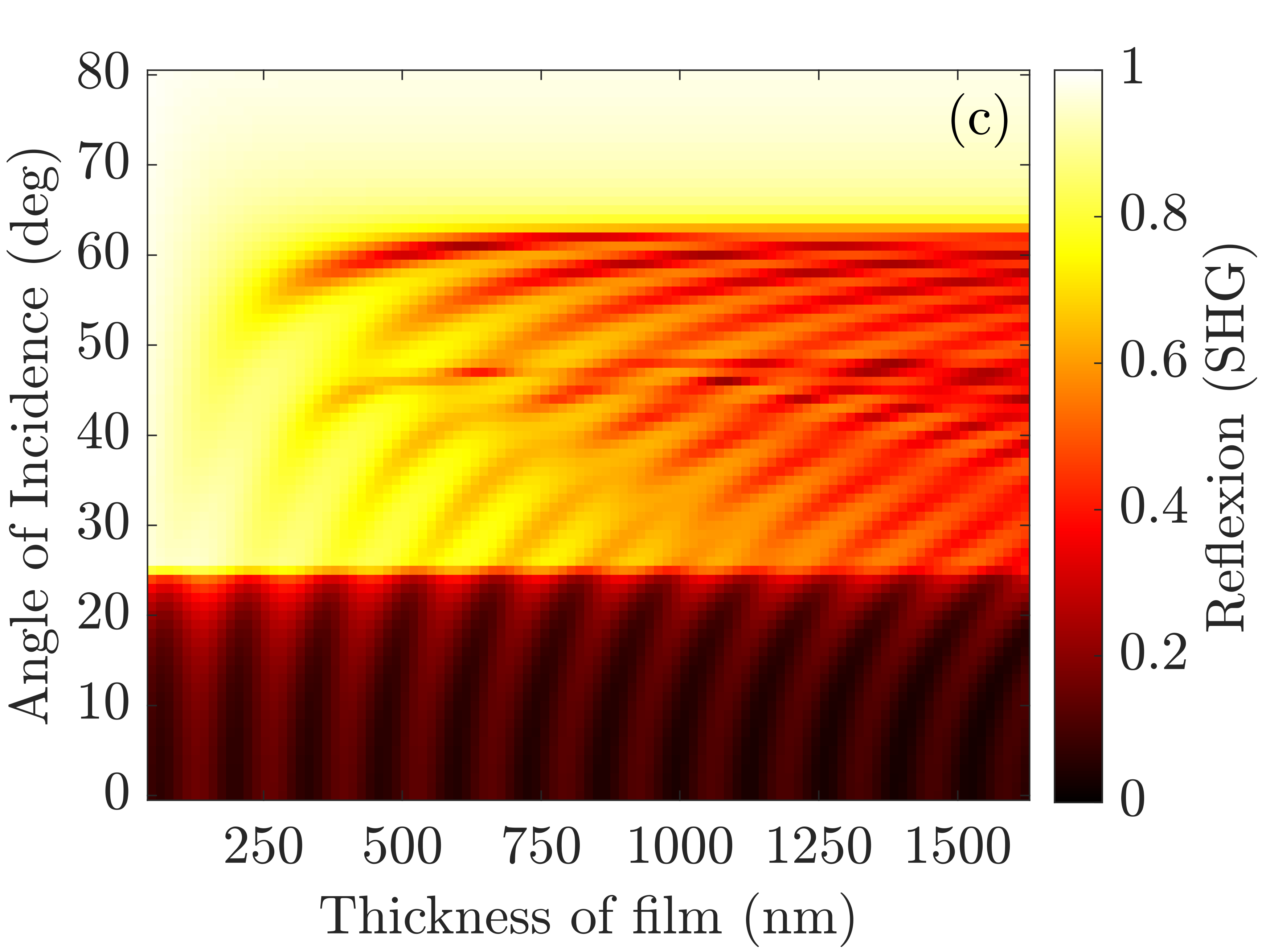}
}
	\caption{\textbf{(a)} Setup of the molecular film on top of a Ceria substrate. The film is illuminated with a plane wave at the fundamental wavelength of 1064\,nm through the substrate under oblique incidence. The light is reflected at the interface between the substrate and the film and between the film and vacuum. In the subsequent analysis, also a total reflection at the second interface is considered. Note that due to the assumption that the substrate is dispersionless, the angle of propagation of both the fundamental and the SHG wave propagating in the negative z-direction in the substrate is the same. \textbf{(b)} Reflection of an incident plane wave at the fundamental and \textbf{(c)} at the SHG frequency. There are two regions of angles of incidence for which the reflection changes rapidly, corresponding to total reflection at the two different interfaces of the system. Between those two regions, Fabry-Perot modes occur.}
    \label{fig:LinearAngles}
	\end{figure*}

In the following, we analyze a more realistic situation in which the film is illuminated through a substrate. The setup is depicted in Figure~\ref{fig:LinearAngles}\textbf{(a)}. Simultaneously, we aim at enhancing the SHG signal as the nonlinear signal originating from the surface of a molecular film is usually very small. For this purpose, we vary the angle of incidence of the fundamental TE-polarized plane wave in the xz-plane. For the material of the substrate, we choose Ceria (CeO$_2$). According to \cite{KANAKARAJU1997191}, the refractive index of Ceria is approximately $n_{\mathrm{subs}}=2.36$ at a wavelength of 600\,nm. We choose this refractive index for both the fundamental and the SHG frequency. With the homogenization approach from \cite{https://doi.org/10.1002/adom.202201564}, we compute a refractive index of the molecular film for the y-axis of approximately $n_{\mathrm{film,yy,fund}}=1.79$ for the fundamental frequency and $n_{\mathrm{film,yy,SHG}}=1.88$ for the SHG frequency. As the refractive index of the substrate is the largest one in the system, two possible total reflections are occuring at the interface between the molecular film and vacuum at an angle of incidence of $\theta_{\mathrm{c},1}=\arcsin{(1/n_{\mathrm{subs}})}=25^{\circ}$ and at the interface between the substrate and the molecular film at an angle of incidence of $\theta_{\mathrm{c},2,\mathrm{fund}}=\arcsin{(n_{\mathrm{film,yy,fund}}/n_{\mathrm{subs}})}=49^{\circ}$ for the fundamental frequency and $\theta_{\mathrm{c},2,\mathrm{SHG}}=\arcsin{(n_{\mathrm{film,yy,SHG}}/n_{\mathrm{subs}})}=53^{\circ}$ for the SHG frequency. 

Figures~\ref{fig:LinearAngles}\textbf{(b)} and \textbf{(c)} show the reflection of an incident plane wave with different angles of incidence at the fundamental and the SHG frequency, respectively. We observe two regions where the reflection rapidly increases. These regions correspond to the angles of incidence for which total reflection occurs at the different interfaces. As the anisotropy of the film is neglected in the calculation of the angle of incidence, there is a small deviation from $\theta_{\mathrm{c},2,\mathrm{SHG}}$. Between these two angles, one observes Fabry-Perot modes corresponding to the reflection from both interfaces.    

\begin{figure*}[]
\centering
     \subfloat{
	\includegraphics[width=0.47\textwidth]{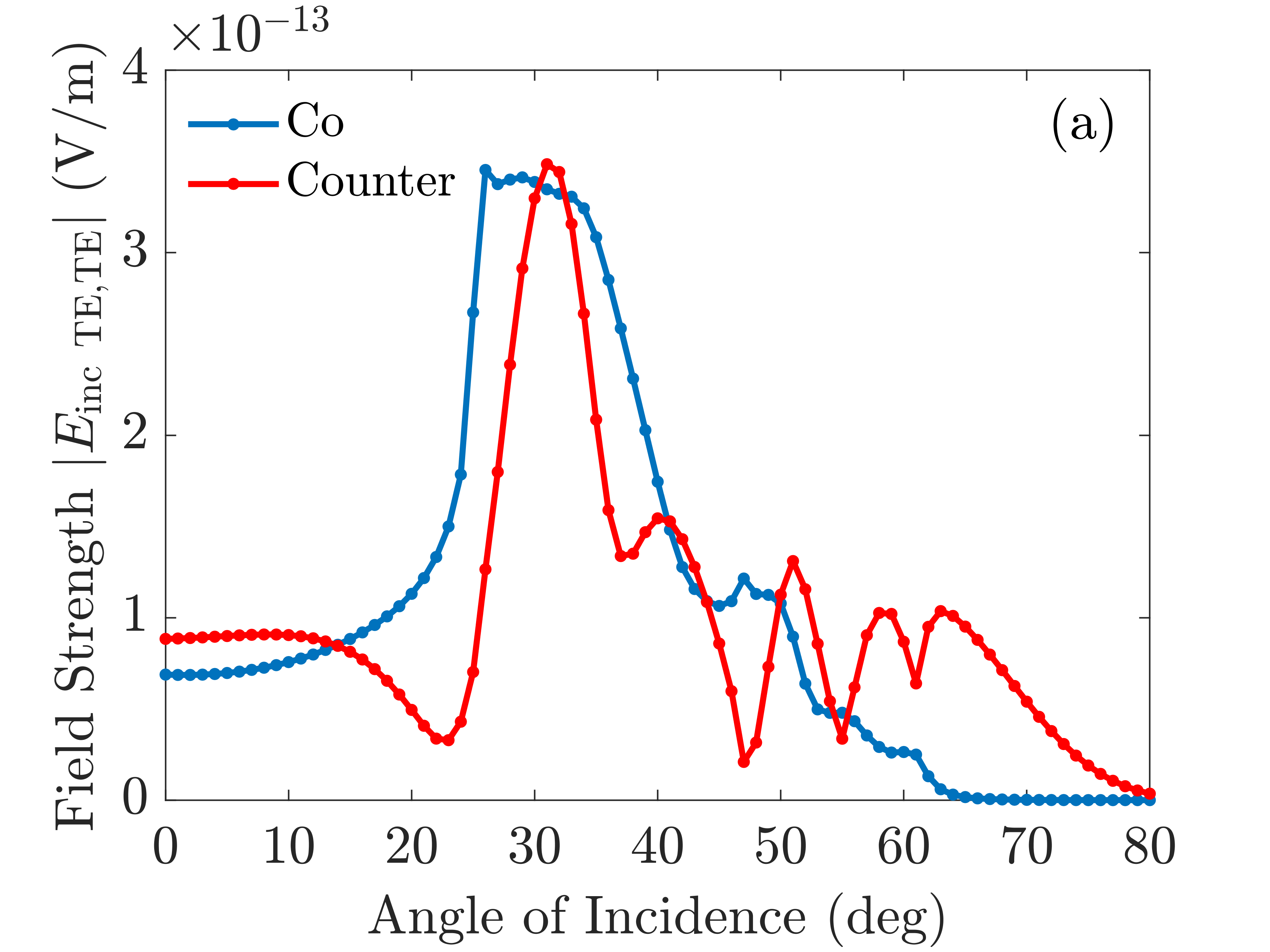}
 }\hspace{0cm}
 \subfloat{
	 \includegraphics[width=0.47\textwidth]{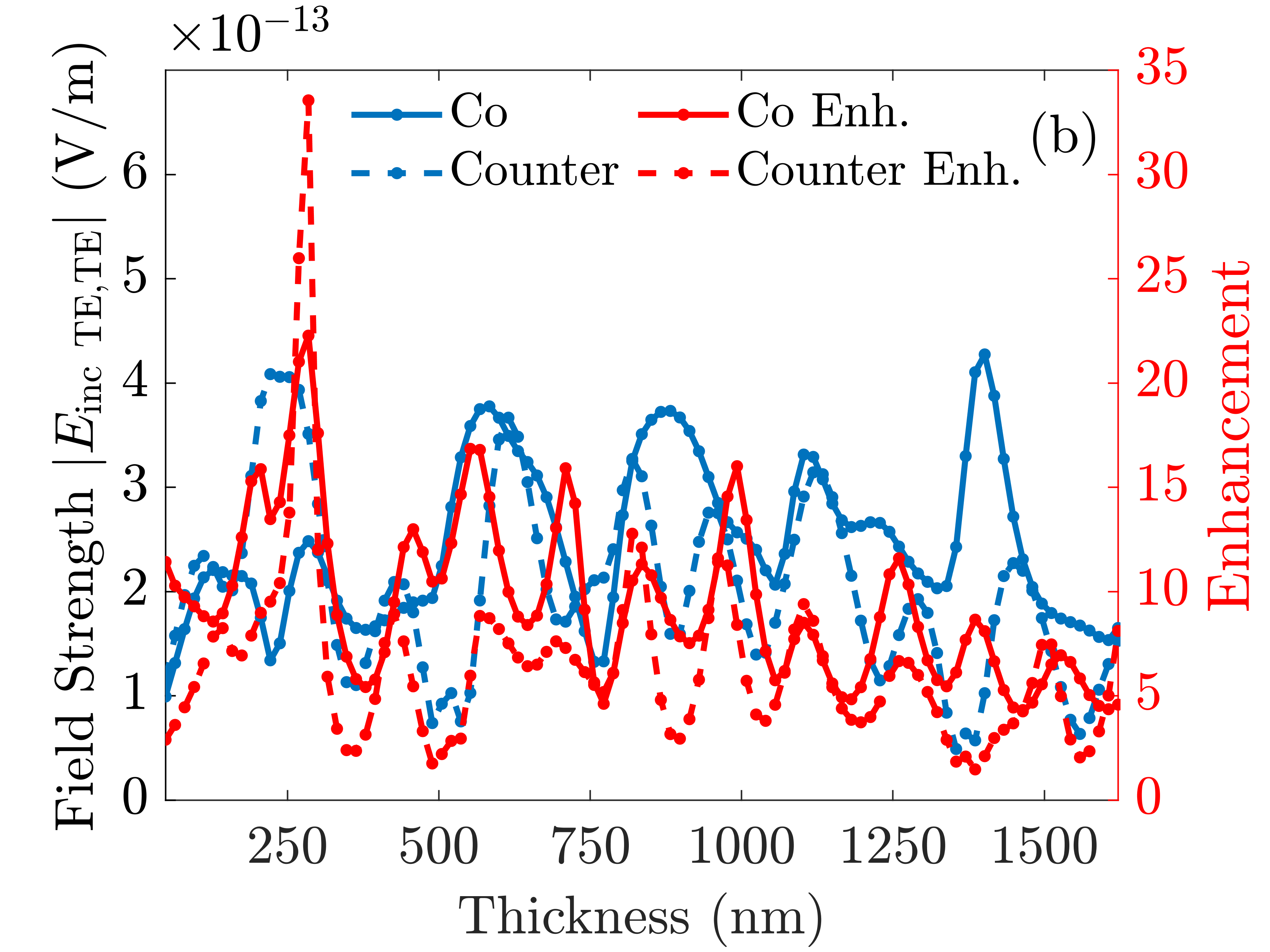}
}
	\caption{\textbf{(a)} SHG fields of the molecular film of a thickness of 631\,nm for different angles of incidence of the fundamental wave. The counterpropagating SHG wave shows a maximum of the field strength for an angle of incidence of $31^{\circ}$. \textbf{(b)} SHG fields for an angle of incidence of $31^{\circ}$ for different values of the thickness of the film. The maximum enhancement of the counterpropagating SHG field strength is 33, which corresponds to an enhancement of the intensity by a factor of 1089.}
    \label{fig:SHGOblique}
	\end{figure*}

To determine the most suitable angle of incidence to enhance the SHG signal compared to the SHG response of the thin film in a vacuum for normal incident illumination, we choose a thickness of the film of 631\,nm. A film of that thickness corresponds to 371 bulk layers. The thickness was chosen because the SHG fields co- and counterpropagating to the illumination show a maximum at normal incidence in the spectrum in Figure~\ref{fig:LayersNormalIncidence}\textbf{(b)}. We compute from that layer stack the SHG response for different angles of incidence of the fundamental wave. Figure~\ref{fig:SHGOblique}\textbf{(a)} shows the spectra of the SHG fields at the interface between film and vacuum (Co) and in the substrate (Counter) for different angles of incidence. We set the angle of incidence to $31^{\circ}$, for which the SHG field, which propagates downward and which shall be detected, shows a maximum in comparison to the SHG response of the thin film in vacuum for normal incident illumination. For this angle of incidence, the structure shows total reflection at the interface between film and vacuum so that the z-component of the wave vector of the SHG field leaving the structure in the upward direction is imaginary. Additionally, the structure shows Fabry-Perot modes in this regime so that the interaction between the light and the structure is enhanced, and constructive interference between the SHG fields originating from the top and bottom surface layer is possible. It is known that Fabry-Perot modes can enhance an SHG signal \cite{doi:10.1021/acs.nanolett.1c03824,doi:10.1021/acsnano.2c03033,PhysRevA.99.043844,PhysRevA.104.063502,PhysRevLett.127.153901,first_paper}. The total reflection at the top surface is leading to a better finesse of the Fabry-Perot modes.

In Figure~\ref{fig:SHGOblique}\textbf{(b)}, the spectra of the SHG fields at the interface between film and vacuum (Co) and in the substrate (Counter) are shown for different values of the thickness of the film for an angle of incidence of the fundamental wave of $31^{\circ}$. We observe a maximum enhancement of 33 for the counterpropagating SHG wave for a thickness of 285\,nm corresponding to 151 bulk layers. The intensity would be enhanced by a factor of $33^2=1089$ in this case. The enhanced field strength is comparable to the values obtained in simulations of a film of a non-centrosymmetric urea molecular crystal \cite{first_paper}, which has also been studied experimentally \cite{Miniewicz2019}.

With this multiscale demonstration, we have shown how the surface nonlinearity of a finite molecular structure translates into the measurable nonlinear response of a molecular film. Additionally, we have considered and designed a simple device to enhance the SHG field by a factor of 33 and its intensity by 1089.

In conclusion, we have introduced a theoretical and computational framework to determine the surface second harmonic generation of centrosymmetric molecular materials. Our framework starts from \emph{ab initio} calculations at the molecular level and then reaches the macroscopic scale for predicting optical SHG measurements. The microscopic origin of surface SHG is identified by layer-dependent models with embedding partial charges combined with DFT calculations. These simulations clearly show increasing symmetry-breaking distortions of the electron cloud around the molecules as the surface layer is approached. The corresponding TD-DFT calculations of the first hyperpolarizabilities show second harmonic generation at the molecular level due to such symmetry-breaking. We have used the methodology to address long-lasting questions about the emergence of the surface SHG signal of crystalline materials. We show that, in a molecular crystalline film with a centrosymmetric unit cell, the SHG is mainly generated at the surface layer: the intensity of SHG at the surface layer is two orders of magnitude larger than at the next layer below and three orders of magnitude larger than two layers below. 

Furthermore, the combination of T-matrices and hyper-T-matrices in this scale-bridging approach opens a novel and efficient path not only to study but also to design and optimize new photonic devices and new nonlinear materials \emph{in silico} before they reach the money- and time-consuming stage of their fabrication and characterization. For example, we show that a relatively simple Kretschmann-like setup can enhance a thousand-fold the surface SHG of a crystalline film of centrosymmetric molecular unit cells. Finally, the methodology can be extended to interfaces between two different materials, as well as to other nonlinear effects such as sum-frequency generation, second harmonic generation circular dichroism\cite{Petrov} and two-photon absorption. Generally, our approach allow us complementary theroretical insight into nonlinear process of not only thin films but also experimentally realized layered materials of same or different molecular composition, metasurfaces\cite{Huttunen2018} or nanostructures\cite{Hecht2023} as ones reported by Hsu \emph{et al.}\cite{doi:10.1021/nn500228r}, Qin \emph{et al.}\cite{doi:10.1021/acsnano.8b06308}, Vabishchevich \emph{et al.}\cite{Vabishchevich} or Butet \emph{et al.}\cite{doi:10.1021/acsnano.5b04373} We foresee very fruitful applications of the presented methodology, in particular for the fast-growing exploitation of nonlinear optical effects in medicine, telecommunications\cite{Celebrano2015, Celebrano2017, Hecht2022}, computing, imaging\cite{Huttunen2013} and sensing.

\section{Methodology}
\subsection{Details of Quantum Chemistry Calculations}

In this work, we focused on two different materials, the non-centrosymmetric urea and the centrosymmetric 7,9-Dibromobenzo[\emph{h}]quinolin-10-ol molecular crystals. The natural population analysis (often also called Natural bond orbital analysis (NBO), in TURBOMOLE, the keywords to invoke this calculation are "\$pop nbo") of the partial charges of the urea used for the embedding of the quantum region were calculated with a developer version of the TURBOMOLE electronic structure program package \cite{TURBOMOLE2022} based on the density functional theory (DFT). We used the hybrid PBE0 exchange-correlation (XC) functional \cite{adamoReliableDensityFunctional1999, ernzerhofAssessmentPerdewBurke1999} and def2-TZVP basis set \cite{weigendBalancedBasisSets2005, weigendAccurateCoulombfittingBasis2006}. The values of the partial charges are presented in the Table~S1. The complex dynamic electric-electric, electric-magnetic, and magnetic-magnetic polarizability tensors in the spectral ranges from 964-1164\,nm (step of 2\,nm) and 482-582\,nm (step of 1\,nm) were calculated using time-dependent density functional theory (TD-DFT) with the same XC functional and basis set. The damping was set to 0.05 eV at the half-width at half-maximum (HWHM). Complex first hyperpolarizabilities with the same damping were calculated using the quantum chemistry package DALTON2020.1 \cite{DALTON, DALTON2}, as it was the only DFT code available at that time to offer such calculations. Calculations of the first hyperpolarizabilities for the molecules and hybrid organo-metallic nanoclusters have been demonstrated previously.\cite{Sanader, Sanader2} To show the equivalency and interchangeability of DALTON and TURBOMOLE codes, we calculated the 50 lowest discrete electronic transitions for the one-photon absorption spectrum in both programs and with both the bulk and surface embedding models, as presented in Figure~S1. As expected, they were equal, which shows that we can combine the results from both programs if necessary due to the different capabilities to calculate linear and nonlinear optical properties. Nevertheless, we had to switch the XC functional to the, similar quality, hybrid B3LYP functional \cite{Becke3P,LYP} since DALTON does not support the chosen PBE0 functional for the quadratic response calculations. Another limiting factor of the DALTON program is the fact that it does not support the resolution-of-identity algorithm \cite{eichkornAuxiliaryBasisSets1995, eichkornAuxiliaryBasisSets1997} for the DFT calculations, thus rendering it to be rather slow for calculating the quadratic response for systems of over $\sim$100 atoms efficiently. 

To overcome all limitations of DALTON and to address much larger systems of interest, we extended the existing TURBOMOLE implementation of real first hyperpolarizabilities to complex frequencies \cite{TM_today_tomorrow,first_paper}. All further calculations for the much larger model of centrosymmetric 7,9-Dibromobenzo[\emph{h}]quinolin-10-ol crystalline material were performed by the TURBOMOLE code. The quantum region comprised four unit cells stacked in the x-direction, totaling 384 atoms. The periodic vectors of our four unit cell model are: \emph{a} = 15.828 \AA, \emph{b} = 18.042 \AA, ~and \emph{c} = 15.821 \AA, with angles between the vectors: $\alpha$ = $\gamma$ = 90° and $\beta$ = 96.139°. The same PBE0 functional and def2-TZVP basis set were used. To speed up the calculations, we combined the resolution-of-identity (RI) with the multipole accelerated resolution-of-identity (marij) \cite{sierkaFastEvaluationCoulomb2003} algorithm, as well as the semi-numerical (senex, esenex) \cite{Holzer2020senex} approach for the exchange, effectively giving us over hundred times speed-up in the calculation over the DALTON program while maintaining the same quality of the results. Again, NBO partial charges calculated and presented in Table S2 for the single unit cell of the material were used to construct the embedding field. In these models, the embedding field had 106,752 point charges and covered the volume of 3x3x31 periodic cells with vectors and angles defined above. 

All TD-DFT calculations were uploaded to the NOMAD database under the following DOI: \url{https://doi.org/10.17172/NOMAD/2023.10.30-1} and to Radar4KIT at the following DOI: \url{https://doi.org/10.35097/1783}.
All other datasets generated and/or analyzed during the current study are available from the corresponding authors on reasonable request.

\subsection{Details of the Optical Scattering Simulations}
In this work, we use the T-matrix-based workflow for nonlinear optical simulations of macroscopic devices presented in \cite{first_paper}. In the first step, the first hyperpolarizability tensor is computed as described in the previous subsection. In the second step, the Hyper-T-matrix $\mathbf{T}^{\mathrm{Hyper}}(-\Omega;\omega,\omega)$ is calculated from the first hyperpolarizability tensor $\mathbf{\beta}(-\Omega;\omega,\omega)$ with Equation~(5) from \cite{first_paper}
\begin{align}\label{eq:HyperTHyperPol}
\begin{split}
    -6\pi\mathbf{C}^{-1}\sum_{m,r,s}\bm{\hat{e}}_{m}\mathrm{T}^{\mathrm{Hyper}}_{mrs}(-\Omega;\omega,\omega)\bm{\hat{e}}_{r}^{\dagger}\cdot \bm{E}_{1}(\omega)\bm{\hat{e}}_{s}^{\dagger}\cdot \bm{E}_{2}(\omega) \\ = \frac{c_{\mathrm{h}}(\Omega)Z_{\mathrm{h}}(\Omega)(k_{\mathrm{h}}\left(\Omega\right))^3}{2\sqrt{6\pi}}\sum_{i,j,k}\bm{\hat{e}}_{i}\beta_{ijk}(-\Omega;\omega,\omega)E_{1j}(\omega)E_{2k}(\omega),
    \end{split}
\end{align}
where $\mathbf{C}^{-1}$ transforms a scattering vector at the SHG frequency from the spherical to the Cartesian basis \cite{Fernandez-Corbaton:2020}. $\bm{\hat{e}}_{m,r,s}$ are spherical basis vectors with the angular momentum numbers $m,r,s$. $\dagger$ denotes complex conjugation and transposition. $\bm{\hat{e}}_{i}$ is a Cartesian basis vector. $\bm{E}_{1}(\omega)$ and $\bm{E}_{2}(\omega)$ are the fields of the fundamental waves, which cancel out while solving Equation~(\ref{eq:HyperTHyperPol}). $c_{\mathrm{h}}(\Omega)=1/\sqrt{\varepsilon_{\mathrm{h}}(\Omega)\mu_{\mathrm{h}}(\Omega)}$ is the speed of light in the medium surrounding the molecules, $Z_{\mathrm{h}}(\Omega)=\sqrt{\mu_{\mathrm{h}}(\Omega)/\varepsilon_{\mathrm{h}}(\Omega)}$ is the wave impedance of the host medium, and $k_{\mathrm{h}}(\Omega)=\Omega\sqrt{\varepsilon_{\mathrm{h}}(\Omega)\mu_{\mathrm{h}}(\Omega)}$ is its wave number. $\epsilon_{\mathrm{h}}(\Omega)$ is the permittivity and $\mu_{\mathrm{h}}(\Omega)$ is the permeability of the medium surrounding the molecules. $\omega$ is the frequency of the fundamental waves, $\Omega$ is the frequency of the SHG wave. 

The Hyper-T-matrix is used to compute the multipolar expansion coefficients $c_{1m}^{\Omega}$ of the SHG scattering response of a finite molecular structure with Equation~(3) from \cite{first_paper}
\begin{align}\label{eq:RelScatHyper}
    c_{1m}^{\Omega}
    = \sum_{r,s} \mathrm{T}^{\mathrm{Hyper}}_{mrs}(-\Omega;\omega,\omega)a_{1r}^{\omega}a_{2s}^{\omega}\mathrm{.}
\end{align}
$a_{1r}^{\omega}$ and $a_{2s}^{\omega}$ are the multipolar expansion coefficients of the incident fundamental waves, which can be computed with Equation~(4) of \cite{first_paper}. The SHG scattering response of a two-dimensional lattice can be obtained with Equation~(6) from \cite{first_paper}
\begin{align}
    \bm{c}^{\Omega}_{\mathrm{tot}} = \left(\mathds{1}-\mathbf{T}(\Omega,\Omega)\sum_{\bm{R}\neq 0}\mathbf{C}^{(3)}(-\bm{R})\mathrm{e}^{\mathrm{i}\bm{k}_{\parallel}(\Omega)\bm{R}}\right)^{-1}\bm{c}^{\Omega}\mathrm{,}
\end{align}
where $\mathbf{T}(\Omega,\Omega)$ is Waterman's T-matrix for linear optical processes at the SHG frequency, which can be computed from the linear polarizabilities of the molecular structure \cite{Fernandez-Corbaton:2020}. $\bm{R}$ are lattice vectors. $\mathbf{C}^{(3)}(-\bm{R})$ contains the translation coefficients of vector spherical waves. $\bm{k}_{\parallel}(\Omega)$ is the component of the wave vector of the zeroth diffraction order of the scattered wave at the SHG frequency, which is parallel to the two-dimensional lattice. With Equation~(21) from \cite{Beutel:21} and the Q-matrices from Equations~(6)-(9) and (12a,b) from \cite{Beutel:21} for the fields at the SHG frequency, we compute the SHG response of stacked molecular lattices in combination with substrates and claddings.

Concerning the workflow for the simulations of the linear response of the molecular structures, we refer to our previous work \cite{SURMOFCavity,https://doi.org/10.1002/adfm.202301093}.

%\begin{acknowledgement}
\section{Acknowledgement}

M.K. and C.R. acknowledge support by the Deutsche Forschungsgemeinschaft (DFG, German Research Foundation) under Germany’s Excellence Strategy via the Excellence Cluster 3D Matter Made to Order (EXC-2082/1-390761711) and from the Carl Zeiss Foundation via the CZF-Focus@HEiKA Program. M.K., C.H., and C.R. acknowledge funding by the Volkswagen Foundation. I.F.C. and C.R. acknowledge support by the Helmholtz Association via the Helmholtz program “Materials Systems Engineering” (MSE). B.Z. and C.R. acknowledge support by the KIT through the “Virtual Materials Design” (VIRTMAT) project. The authors thank Dr. Markus Nyman for the fruitful discussions. M.K. and C.R. acknowledge support by the state of Baden-Württemberg through bwHPC and the German Research Foundation (DFG) through grant no. INST 40/575-1 FUGG (JUSTUS 2 cluster) and the HoreKa supercomputer funded by the Ministry of Science, Research and the Arts Baden-Württemberg and by the Federal Ministry of Education and Research.

%\end{acknowledgement}

%\begin{suppinfo}
%\section{Supporting information}

%The following files are available free of charge.
%\begin{itemize}
%  \item Filename: supplementary\_information.pdf
%\end{itemize}

%\end{suppinfo}

\bibliography{bibliographyArxiv}

\end{document}